# Blockchain Framework for Artificial Intelligence Computation

Jie You[1,2,*]

[1]Dasudian Technologies Ltd., Shenzhen, 518057, China
[2]Institute of Computer Engineering, Heidelberg University, Heidelberg, 69117, Germany
[*]barco@dasudian.com

## Abstract

Blockchain is an essentially distributed database recording all transactions or digital events among participating parties. Each transaction in the records is approved and verified by consensus of the participants in the system that requires solving a hard mathematical puzzle, which is known as proof-of-work. To make the approved records immutable, the mathematical puzzle is not trivial to solve and therefore consumes substantial computing resources. However, it is energy-wasteful to have many computational nodes installed in the blockchain competing to approve the records by just solving a meaningless puzzle. Here, we pose proof-of-work as a reinforcement-learning problem by modeling the blockchain growing as a Markov decision process, in which a learning agent makes an optimal decision over the environment's state, whereas a new block is added and verified. Specifically, we design the block verification and consensus mechanism as a deep reinforcement-learning iteration process. As a result, our method utilizes the determination of state transition and the randomness of action selection of a Markov decision process, as well as the computational complexity of a deep neural network, collectively to make the blocks not easy to recompute and to preserve the order of transactions, while the blockchain nodes are exploited to train the same deep neural network with different data samples (state-action pairs) in parallel, allowing the model to experience multiple episodes across computing nodes but at one time. Our method is used to design the next generation of public blockchain networks, which has the potential not only to spare computational resources for industrial applications but also to encourage data sharing and AI model design for common problems.

## Introduction

Since the appearance of Bitcoin[1], blockchain technologies have brought about disruptions to traditional business processes[2,3,4], have been used for industrial advance[5-11], and have even triggered innovations in biotech and medical applications[12-16].

Blockchain seeks to minimize the role of trust in achieving consensus[2]. There are different consensus mechanisms exit[17], where the most well-known is the proof-of-work that requires solving a complicated computational process, such as finding hashes with specific patterns. This consensus algorithm disincentivizes misbehavior by making it costly for any agent to alter the state, so there is no need for trust in any particular central entity. Although there are other mechanisms for achieving consensus, proof-of-work is self-sufficient and rent-free simultaneously[18].

Proof-of-work systems have several major benefits. First, they are an excellent way to deter spammers. In addition, proof-of-work systems can be used to provide security to an entire network. If enough nodes

(computers or dedicated mining machines) compete to find a specific solution, then the computational power needed to overpower and manipulate a network becomes unattainable for any single bad actor or even a single group of bad actors.

However, there is a primary disadvantage to proof-of-work systems. They consume a large amount of computing power and waste energy, as additional electricity is used for computers to perform extra computational work. This can add up to an extremely large amount of excess electricity consumption and environmental detriment[19,20,21].

Machine-learning technology has been powering many aspects of modern society, from web searches to content filtering on social networks to recommendations on e-commerce websites, and it is increasingly present in consumer products such as cameras and smartphones. Machine-learning systems are used to identify objects in images[22], transcribe speech into text[23], match news items, posts or products with users' interests, and select relevant search results. Particularly with the boom in digital data on the Internet, deep learning, as a representation-learning method, has shown great power in driving myriad intelligent applications and will have many more successes in the near future[24]. Because it requires very little engineering by hand, deep learning can easily take advantage of increases in the amount of available computation and data[24].

As one branch of machine learning technology, reinforcement learning is the task of learning what actions to take, given a certain situation or environment, to maximize a reward signal. In contrast to deep learning, which is a supervised process, reinforcement learning uses the reward signal to determine whether the action (or input) that the agent takes is good or bad. Reinforcement learning has inspired research in both artificial and biological intelligence[25,26] and has been widely used in dynamic task scheduling[27], planning and cognitive control[28], and more interesting topics have been in active research[29].

To use machine learning in practical scenarios, generally plenty of computational power is required to support so-called artificial intelligence (AI) model training and execution at different scales according to the complexity of models and the amount of data to be processed. For instance, GPT-3[30] and Switch Transformers[31] have shown that AI model performance scales as a power law of model size, dataset size and amount of computation. The cost of AI is increasing exponentially to achieve the desired target with a larger model size and more crunched data. In general, when AI models and the training datasets are large enough, the models need to be trained for more than a few epochs to learn fully from the data and generalize well; therefore, the hardware cost and time cost are both high for well-performing AI applications.

On the one hand, blockchain systems waste a large amount of computational power to solve the meaningless puzzles for proof-of-work, and on the other hand, many useful AI applications require substantial computing capacities to achieve high performance. To balance these two aspects, in this paper, we present a blockchain model that combines the computation for proof-of-work and for artificial intelligence model learning procedures as one process, achieving a consensus mechanism of blockchain and artificial intelligence computation simultaneously and in an efficient way.

## The blockchain model

In this paper, we model the blockchain system as an agent of reinforcement learning. As depicted by Fig. 1, every block represents a state of a Markov state machine, whereas the creation and linking process of blocks is a Markov decision process (MDP)[29], with the following setup:

1. The environment is defined as oracle in this blockchain system, which provides the data to blockchain via its state transitions ($S_t \rightarrow S_{t+1}$).
2. In the present state ($S_t$), the agent chooses an action ($A_t$) according to the current **policy** ($\pi_t$) and receives a reward ($R_{t+1}$) from the environment, while the state of the environment transforms from $S_t$ to $S_{t+1}$. Afterwards, the nodes of blockchain train the policy model and update it from $\pi_t$ to $\pi_{t+1}$, which are stored in the memory of computing nodes as the function for choosing the next action by feeding the next state. The computation that occurs in this process is defined as the proof-of-work for the computing nodes, which compete to do so in the blockchain system.
3. Computing nodes of the system create a new block, recording the current state of environment ($S_{t+1}$), the last chosen action ($A_t$), the reward ($R_{t+1}$) received from the environment, the data ($D_{t+1}$) to be written onto blockchain for a transaction, and the Hash value of last block ($h_{t+1} = Hash(S_t, A_{t-1}, R_t, D_t, h_t)$), as shown in Fig. 2. When a node finishes the computation of proof-of-work and creates a new block, it is saying that a mining process is completed.
4. When a mining process completes, the newly created block is linked to the previous block by the hash value of the previous block (Fig. 2).

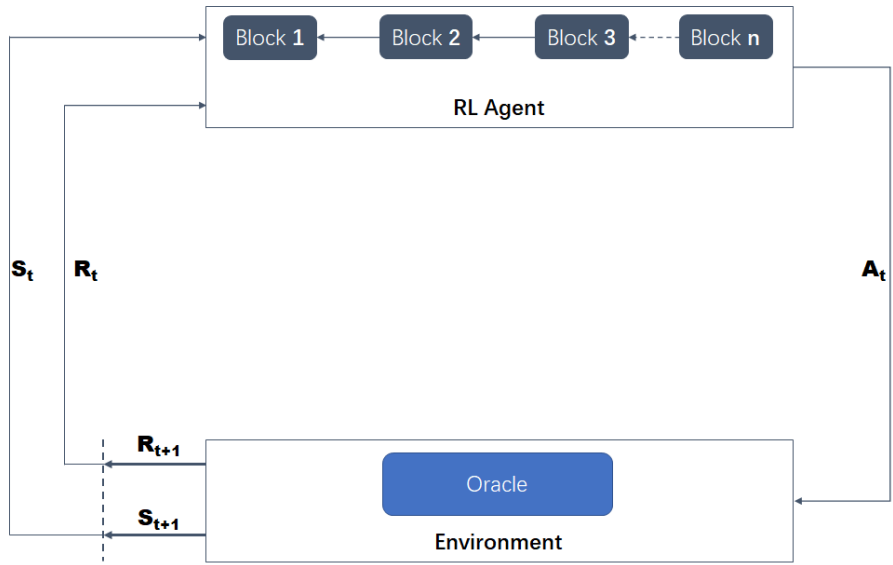

**Figure 1 The blockchain model based on reinforcement learning**

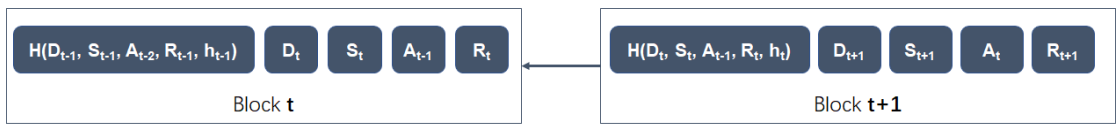

**Figure 2 The mechanism for blocks to store data and being linked**

In any block of the chain, the stored Hash value of the previous block prevents the data from being falsified because if any data are changed, the block's Hash value must be different and in turn change the data stored in the next block, which invalidates the linkage of blocks within the chain. In addition, if the state of the environment ($S_t$) or action ($A_{t-1}$) stored in one block is modified, the next state ($S_{t+1}$), next action ($A_t$) and reward ($R_{t+1}$) will probably be different from those actually stored in the next block when transformed by the **policy**, which also largely decreases the possibility of and increases the difficulty of tampering with data.

## Proof-of-work

The proof-of-work algorithm is implemented as follows:

1. At present state ($S_t$) choose an action ($A_t$) based on current ***policy*** ($\pi_t$);
2. Exert $A_t$ onto the environment, or say interact with the oracle, receiving a reward ($R_{t+1}$), and the state of environment changes to $S_{t+1}$;
3. Based on the state transition ($S_t \rightarrow S_{t+1}$), action selected ($A_t$) and the reward received ($R_{t+1}$), the nodes of blockchain train the predefined action-value function of the reinforcement-learning model and update the ***policy*** to $\pi_{t+1}$.

In this paper, the proof-of-work includes the computing processes of selecting action, generating reward regulated by current ***policy*** ($\pi_t$), and training the action-value function model and updating the ***policy***. Considering many practical MDP problems, the state spaces are large enough or even with unlimited states, which require large and complicated deep neural networks to achieve a well-performing approximator of the action-value function, so the computation of proof-of-work is highly resource-demanding. Therefore, any attempts to tamper with data or hack the whole blockchain are almost unachievable due to the daunting cost of computing resources and time.

## Consensus based on rewarding of reinforcement-learning

When a node working for the blockchain finishes proof-of-work, or say a mining process, it needs to synchronize the newly generated block to other nodes in the network to guarantee the consistency of data within the whole network. However, because of the occurrences of network delay, errors and attacks, nodes may keep different versions of the blockchain information, resulting in inconsistency. Therefore, we design a consensus mechanism for nodes to achieve data consistency across the whole network, as follows:

1. First, prioritize the longest chain: if nodes keep chains of different lengths, then the longest chains should be chosen as the proven chains;
2. If at step 1, there is more than one chain kept, there are two optional ways to determine the final chain:
   a. Comparing the reward value ($R$) at the last block of the chains, choose the chain with the maximum reward as the final consented chain.
   b. Comparing the sum of rewards ($\sum R_t$) across all blocks of the chain, choose the one with the maximum summation as the final consented chain.

Although different nodes share the same ***policy*** algorithm, they experience self-unique model training and ***policy*** updating processes and keep their own action-value function model and ***policy*** instances in memory, which are not synchronized to each other, so for the same state ($S_t$), different nodes will not necessarily select the same action or receive the same reward. This brings about two valuable aspects:

1. Even if more than 51% of the total nodes within the network are hacked, which attempts to falsify the data and regenerate a new chain, when they complete the proof-of-work, the maximum reward ($R_{max}$) is not definitely received by them but rather possibly by the unhacked nodes, in which case the falsified blocks will not be consented. Thus, the consensus mechanism designed in this paper additionally enhances the safety of the blockchain system by reducing the possibility of being hampered.

2. Because every node keeps its own instances of the action-value function model and *policy* and competes to achieve the maximum reward ($R_{max}$) by implementing the proof-of-work, this allows the reinforcement-learning algorithm to learn along more than one path (the number of paths equals the working nodes within the network) on the same environment state and at one time point. It equivalently replaces time with space for AI model training, which achieves multiple epochs of training at one round. In this way, while the blockchain is growing, the reinforcement-learning algorithm backing its proof-of-work and consensus mechanism more fully learns diversified possibilities and converges faster, thereby making more precise prediction ($S_t \rightarrow A_t$) as quick as possible, which is conducive to the overall goal achievement in a shorter term for the reinforcement-learning model. This is specifically beneficial for online learning applications of AI.

In summary, the blockchain system presented in this paper is a distributed training system for reinforcement-learning algorithms, which accelerates the learning process of AI models while realizing blockchain properties.

## Proof-of-work with deep Q-learning

Specifically, we use deep Q-learning[29,32,33] as the *policy* updating algorithm for the agent to learn. The iteration of the action-value function in Q-learning is formulated as:

$$Q(S_t, A_t) \leftarrow Q(S_t, A_t) + \alpha[R_{t+1} + \gamma \boldsymbol{max}_a Q(S_{t+1}, a) - Q(S_t, A_t)] \quad (1)$$

where $Q$ is the action-value function to be learned for the optimal decision; $a$ and $R_{t+1}$ are the selected action and received reward at state $S_{t+1}$, respectively; and $\alpha$ ($0 < \alpha < 1$) and $\gamma$ ($0 < \gamma < 1$) are the step-size parameter and discount-rate parameter, respectively.

A deep neural network is used to represent the $Q$ function, and every node of the blockchain will be the agent to learn the $Q$ function and iterates according to equation (1), with the *policy* determining which state-action pairs are visited and updated.

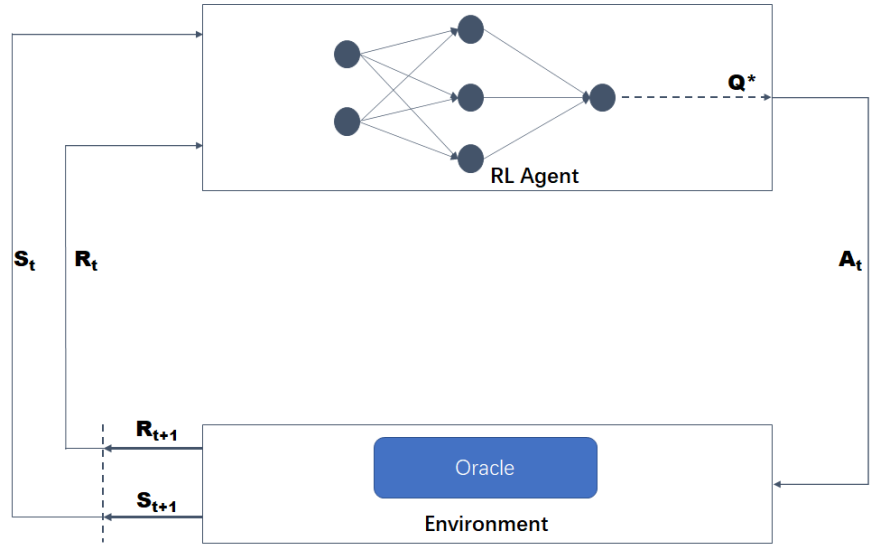

**Figure 3 The blockchain model based on deep Q-learning**

As shown in Fig. 3, at any time step $t$ the nodes of the blockchain calculate the optimal action $A_t$ according to the current $Q$ function and state $S_t$ and then update the $Q$ function according to formula (1) for the next state. Specifically, in this research, we represent the $Q$ function as a deep neural network. As illustrated in Fig. 4, the section in red represents the target, which has the same neural network architecture as the $Q$ function approximator (section in green) but with frozen parameters. For every $C$ iterations (a hyperparameter), the parameters from the prediction network are copied to the target network. A loss function is defined as the mean squared error of the target Q-value and predicted Q-value:

$$Loss = \left(R + \gamma max_a Q(S_{t+1}, a; \theta') - Q(S_t, A_t; \theta)\right)^2 \quad (2)$$

where $\theta'$ and $\theta$ represent the parameters of the target network and prediction network, respectively. Then, this is basically a regression problem, where the prediction network updates its gradient using backpropagation to converge.

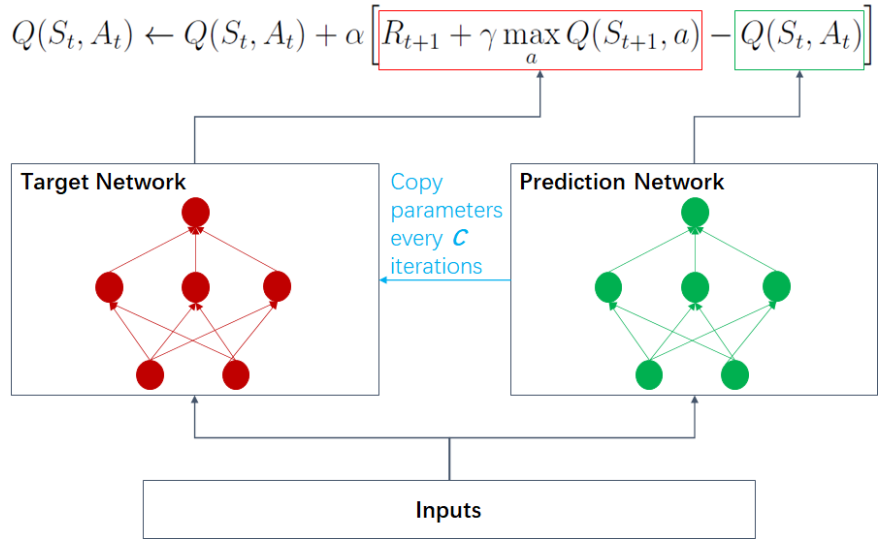

**Figure 4 Schematic diagram for Q function iteration and its neural network representations**

The steps involved in the deep Q-learning procedure for every node of the blockchain are as follows:

1. At time step $t$, every node feeds state $S_t$ into the prediction Q network, which will return the Q-values of all possible actions in the state.

2. Selects an action using an epsilon-greedy policy: with probability epsilon ($0 < \varepsilon < 1$) to select a random action and with probability $1 - \varepsilon$ to select an action that has a maximum Q-value, such as $argmax(Q(S_t, a; \theta))$.

3. Performs this action $A_t$ in state $S_t$ and moves to a new state $S_{t+1}$ to receive reward $R_{t+1}$. Writes this transition information into a new block and stores it in a replay buffer of the node as $(S_t, A_t, R_{t+1}, S_{t+1})$.

4. Next, samples some random batches of transitions from the replay buffer and calculates the loss defined by equation (2).

5. Gradient descent is performed with respect to the prediction network parameters to minimize this loss. Then, the node finishes once proof-of-work computation and proves a newly generated block.

6. After every $C$ iterations, copies the predicted Q network weights to the target network weights.

7. Repeat above steps.

## The awarding mechanism for mining

In this framework, the computations for the reinforcement-learning algorithm and particularly for the training of deep neural networks are assigned to the nodes (mining machines) of blockchain to compete for the proof-of-work, and after nodes complete the proof-of-work, the nodes that are fastest to finish the computation and receive the maximum reward can finally win to prove the blocks, which is the consensus mechanism of this blockchain. Thus, in our design, we stipulate the maximum reward $R_{max}$ as the award to the node that finally wins the competition of proof-of-work and consensus to encourage more computers with better capacity to join the blockchain network and contribute to artificial intelligence computations. This award value $R_{max}$ is called the ***token*** of this blockchain.

## Conclusion

In this paper, we present a blockchain framework that organically stitches computations for reinforcement-learning and proof-of-work as well as a consensus mechanism, achieving a versatile distributed computing system. On the one hand, taking advantage of the complexity and high computing cost of the reinforcement-learning process and deep neural network training increases the difficulty of hacking the blockchain network or falsifying the data. In particular, because the nodes keep self-owned instances of ***policy*** and neural networks, they keep uncertainties of state transition ($S_t \rightarrow S_{t+1}$) and action selection that may be different nodes from nodes. These uncertainties additionally consolidate the stability of chain linkages that are difficult for hackers to mutate. The consensus mechanism of maximum-reward-win adds an additional barrier deterring hackers to tamper with the chain. On the other hand, utilizing the nodes within the blockchain network to fulfil the training and running of AI algorithms naturally contributes computing power to practical intelligent applications. Meanwhile, by distributing the AI model training to multiple nodes that simultaneously crunch the same data generated by the environment, or saying oracle in this blockchain system, the nodes keep their own instances of the AI model, so the nodes experience different paths of learning with different parameter values and hidden states of the AI model at every time step. This equivalently implements multiple epochs of training within only one round of the learning process, which improves the training efficiency and accelerates the convergence of models.

## Discussion

The blockchain framework presented in this paper paves an avenue for AI applications that require intensive computing power and a quicker generalization rate and a credible network for feeding data to AI models. Therefore, this provides a potential solution for facilitating the development of industrial intelligence, which has been developing slowly due to a lack of data, because enterprises in industrial verticals are not willing to share their assets. In addition, in industry, there are either insufficient professional AI talent or computing capacities for AI applications, so this blockchain framework could provide an open platform encouraging AI professionals to contribute their expertise as well as computing resources supporting the advancement of industry. Furthermore, this framework is particularly pragmatic for nonepisodic reinforcement-learning problems with models continuously adapting to the environment, such as financial markets, IoT networks and factory operations.

Ultimately, it could be expected that by combining blockchain and artificial intelligence into one computational framework, the two most important resources, data and computing power, can be utilized in a mutually supportive way over a creditable platform that encourages more innovations in artificial intelligence applications. Finally, we believe that this blockchain framework for AI computation could be a potential backbone of the industrial Internet.